\documentclass{article}
\usepackage{spconf,amsmath,graphicx}
\usepackage{amsmath,amssymb,graphicx,url,times}
\usepackage{multirow}
\usepackage{siunitx}
\usepackage{pifont}
\usepackage{booktabs}
\usepackage{url}
\usepackage[table]{xcolor}
\usepackage{hyperref}

\usepackage{algorithm}
\usepackage{algorithmic}

\title{Unsupervised Source Separation By Steering Pretrained Music Models}
\name{Ethan Manilow$^1$, Patrick O'Reilly$^1$, Prem Seetharaman$^2$, Bryan Pardo$^1$}
\address{$^1$Northwestern University \quad $^2$Descript, Inc.}
\begin{document}
\ninept
%
\maketitle
\begin{abstract}
We showcase an unsupervised method that repurposes deep models trained for music generation and music tagging for audio source separation, without any retraining. An audio generation model is conditioned on an input mixture, producing a latent encoding of the audio used to generate audio. This generated audio is fed to a pretrained music tagger that creates source labels. The cross-entropy loss between the tag distribution for the generated audio and a predefined distribution for an isolated source
is used to guide gradient \textit{ascent} in the (unchanging) latent space of the  generative model. This system does \textit{not} update the weights of the generative model \textit{or} the tagger, and only relies on moving through the generative model's latent space to produce separated sources. We use OpenAI's \textsc{Jukebox} as the pretrained generative model, and we couple it with four kinds of pretrained music taggers (two architectures and two tagging datasets). Experimental results on two source separation datasets, show this approach can produce separation estimates for a wider variety of sources than any tested supervised or unsupervised system. This work points to the vast and heretofore untapped potential of large pretrained music models for audio-to-audio tasks like source separation.

\end{abstract}
\section{Introduction}
\label{sec:intro}

The research area of Music Information Retrieval (MIR) is constrained by a lack of labeled data sets, which limits our ability to train robust systems and evaluate them well. Specifically, the task of musical source separation has been hindered by a dearth of well-labeled data~\cite{manilow2019cutting}.
This leads to severe shortcoming in terms of the range of instrument source classes that current systems can separate. Many systems, in fact, only separate the four classes  (voice, bass, drums and ``other'') in the widely-used MUSDB18~\cite{musdb18} dataset, making them unsuitable for separating most musical instruments.

Simultaneously, the recent availability of large pretrained models has revolutionized generative and discriminative tasks in the domains of computer vision and natural language processing. The combination of VQGAN~\cite{esser2021taming} and CLIP~\cite{radford2021learning} has captured the attention of many artists, who have been captivated by the system's ability to use natural language to create generative art. Similarly, researchers have shown how to steer large pretrained language models for downstream discriminative tasks either using transfer learning~\cite{raffel2020exploring} or so-called few-shot ``prompt engineering''~\cite{brown2020language}. Recent work has taken this ethos to the MIR domain, leveraging the representations learned by the large training regime of an unsupervised generative music model for downstream MIR tasks, like key detection and music tagging~\cite{castellon2021codified}.

\begin{figure}[t]
    \centering
    \includegraphics[width=\columnwidth]{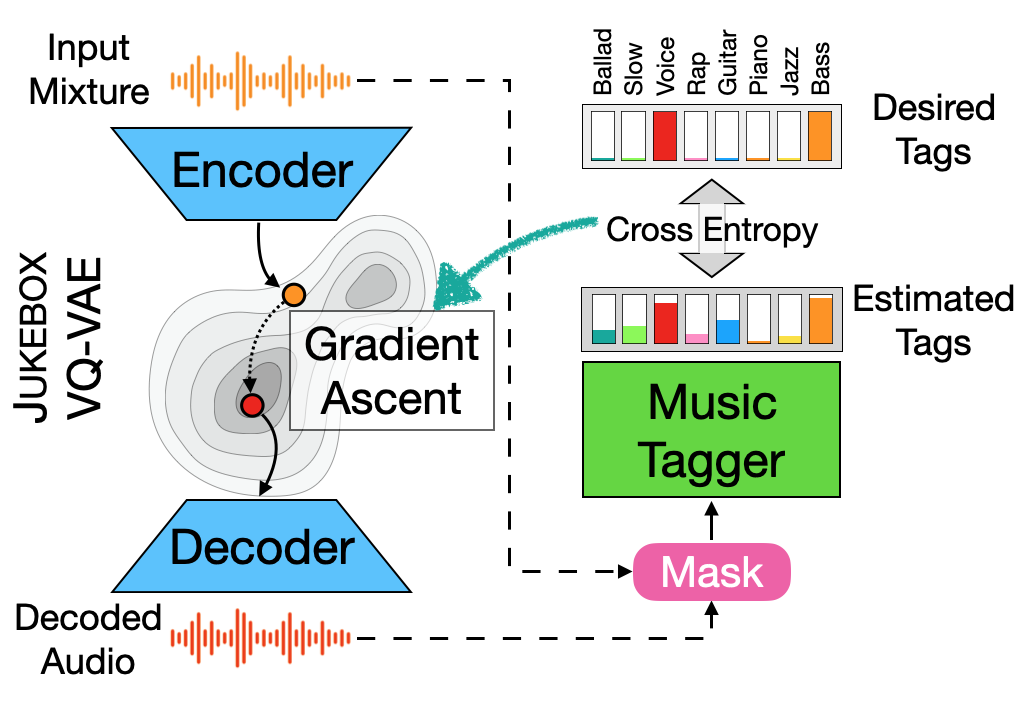}
    \caption{Our system performs gradient ascent in the \textsc{Jukebox} VQ-VAE embedding space such that when the audio is input into a music tagger it matches a predefined set of tags. The weights of VQ-VAE and the Music Tagger are frozen. With this setup we can perform unsupervised source separation.}
    \label{fig:diagram}
\end{figure}

In this work, we further this ethos by exploring how large, pretrained music models can be used for musical source separation, leveraging the vast amounts of unlabeled or weakly labeled data that these models see during training. We combine the VQ-VAE from OpenAI's \textsc{Jukebox}, a generative model of musical audio, with a music tagger. We task \textsc{Jukebox} with producing audio that matches a predefined set of tags that correspond with the musical source we wish to separate. To do this, we perform gradient ascent in the embedding space of the VQ-VAE and use the decoded audio as a mask on the input mixture. We demonstrate experimentally that this setup is able to separate a wider variety of sources than previous purpose-built separation systems consider, all without updating the weights of \textsc{Jukebox} or the tagger. We provide additional demos and runnable code on our demo site.\footnote{\href{https://ethman.github.io/tagbox}{\url{https://ethman.github.io/tagbox}}}


\section{Prior Work}
\label{sec:prior_work}

Recently, many source separation researchers have focused on methods that produce high-quality results on the datasets for which there is sufficient ground truth source data. For instance, the website Papers with Code shows a steady increase in the best performing separation systems on the MUSDB18~\cite{musdb18} dataset over the past few years.\footnote{\href{https://bit.ly/pwc_musdb18}{\url{https://bit.ly/pwc_musdb18}}} Similarly, the recent Music Demixing Challenge~\cite{mitsufuji2021music} invited people to compete to determine the best performing system on a test set that had the same source definitions as MUSDB18. As a result, the community has produced a large number of deep learning-based \textit{supervised} separation systems that are purpose-built to separate sources as defined by MUSDB18. However, the source definitions in MUSDB18 are limiting,~\cite{manilow2019cutting} including isolated source data for only Vocals, Bass, Drums, and a catchall ``Other'' source for all other source types. Furthermore, MUSDB18 is relatively small, totalling 150 songs, which leads the authors of many state-of-the-art systems \cite{mitsufuji2021music, defossez2019music, stoter2019open} to collect additional data  or lean heavily on augmentation. 


Prior to the deep learning era, unsupervised source separation was the norm. One of the most popular algorithms was Non-negative Matrix Factorization (NMF)~\cite{smaragdis1998blind}. While NMF is theoretically flexible enough to separate any source, it often required hand-designed algorithms to determine how to cluster spectral templates into coherent sources. Musical priors, such as repetition~\cite{rafii2012music} or harmonicity vs. percussiveness~\cite{fitzgerald2010harmonic},  have also been used to create unsupervised separation algorithms, however such algorithms are limited to separating sources that match the prior (e.g., a backing band) vs those that do not (e.g., a singing voice), and have been surpassed by deep learning-based methods.

Recent work in speech and environmental sound separation has explored unsupervised deep learning. Mixture Invariant Training (MixIT)~\cite{wisdom2020mixit} is a technique which creates mixtures of mixtures (MoMs) and tasks a network with overseparating each MoM such that when sources are recombined, a mixture reconstruction loss can be used, forgoing the need for isolated source data altogether. While we are unaware of anyone using MixIT for music separation, MixIT makes an implicit presumption that any two sources in a mix are independent~\cite{wisdom2020mixit}, an assumption that may not hold for music. Similarly, Neri et. al~\cite{neri2021unsupervised} propose a technique for training a variational auto-encoder (VAE) for unsupervised source separation, however in our work we do not train networks at all, rather we use frozen, pretrained models for separation.

Previous works have explored using additional networks for separation instead of directly optimizing a separation net on ground truth sources. For instance, the work of Pishdadian et. al.~\cite{pishdadian2020finding} is most similar to ours; they explore using a pretrained sound event detection (SED) system and the goal of the separator network is to maximize estimated SED labels during training. 
Similarly, Hung et. al~\cite{hung2021transcription} use a pretrained transcription network to train a separator. Our work differs from Pishdadian et. al. and Huang et. al. in that we do not \textit{not} train any networks, instead we repurpose off-the-shelf networks that have  \textit{never} been trained for source separation. 

\section{Background}
\label{sec:background}

\subsection{OpenAI's \textsc{Jukebox}}
\label{subsec:jukebox}

OpenAI recently released \textsc{Jukebox}~\cite{dhariwal2020jukebox}, a generative audio model that creates music. \textsc{Jukebox} is composed of two components: a hierarchical VQ-VAE~\cite{razavi2019generating} that learns to turn raw waveforms into tokens and back, and a language model that learns how to generate new tokens which can be passed through the decoder to create musical audio. Both the VQ-VAE and the language model are unsupervised. In this work, we are interested in the VQ-VAE, specifically.

\textsc{Jukebox}'s VQ-VAE is a three-level hierarchical VQ-VAE that generates discrete tokens at different sample rates, compressing the 44.1kHz input audio to tokens with sample rates of 5.51kHz, 1.37kHz, and 344Hz for each level, respectively. Each level has a codebook size of 2048 with each code having 64 dimensions. All levels are trained to reconstruct the input waveform and are optimized with a multi-scale spectral loss. The VQ-VAE also uses a codebook loss to ensure that non-discretized latent vectors are close to their nearest neighbor discretized token vectors and a commitment loss, which stabilizes the encoder. The VQ-VAE is trained on 1.2 million songs scraped from the web. We refer the reader to the \textsc{Jukebox} paper for further training details~\cite{dhariwal2020jukebox}.
Because we are interested in producing the highest-quality separation results possible, we only focus on the ``Bottom'' level, which compresses the input audio to tokens at a sample rate of 5.51kHz.


\begin{algorithm}[t]
\caption{Our Method}

\begin{algorithmic} [1]
\STATE{$e \leftarrow V_{encoder}(x)$ Encode the input mixture.}
\STATE{$X \leftarrow  STFT(x)$}

\REPEAT
\STATE{$j \leftarrow  V_{decoder}(e)$ Decode the embedding.}
\STATE{$j \leftarrow STFT(j)$}
\STATE{$\Bar{M} \leftarrow \frac{|J|}{\max(|J|, |X|) + \varepsilon}$ Build the mask.}
\STATE{$\Bar{S} \leftarrow \Bar{M} \odot X$  Mask the mixture.} 
\STATE{$T_{\Bar{S}} \leftarrow Tagger(ISTFT(\Bar{S}))$ Get the probability over tags.}
\STATE{$e \leftarrow  \delta\nabla L(T_{\Bar{S}}, T_{target};e)$ Update the embedding.}
\UNTIL{max steps}
\STATE{$s_{out} \leftarrow x - ISTFT(\Bar{S})$}
\end{algorithmic}
\end{algorithm}

\subsection{Automatic Music Tagging}
\label{subec:tagging}

Music tagging is the task of labeling musical audio clips with semantic labels called ``tags''~\cite{law2009evaluation, bogdanov2019mtg}. These tags are useful for music search and recommendation systems, enabling automatic labelling of large music corpora. The content that the tags represent can vary, sometimes indicating information about a song's genre, the song's mood or theme, or whether particular instruments are audible. 

Music tagging systems are designed to predict a set of multi-hot, binary labels (i.e., tags) based on the acoustic contents of an input signal. Many recent works use convolutional neural networks at their core, varying the convolutional filter size and input representation of the audio~\cite{won2020eval}. Common datasets for music tagging are an order of magnitude larger than source separation datasets: MagnaTagATune (MTAT)~\cite{law2009evaluation} contains 25,877 30-second labeled audio clips ($\approx$21x more hours of audio than MUSDB18) and MTG-Jamendo (MTG)~\cite{bogdanov2019mtg} contains 55,701 labeled audio clips with a minimum song length of 30 seconds ($\ge$ 46x more hours of audio than MUSDB18). 
We refer the reader to Won et. al. for an overview of recent advances in music tagging~\cite{won2020eval}.

In this work, we use pretrained music taggers provided by Won et. al~\cite{won2020eval}. 
We examine using two pretrained music tagging systems, with each having a different input representation: FCN~\cite{choi2016automatic} with Mel spectrogram inputs, and HarmonicCNN~\cite{won2020data}, which inputs a variant of a constant-Q transform that has learnable filters.
We also explore using taggers trained on different datasets, namely MagnaTagATune (MTAT)~\cite{law2009evaluation} and MTG-Jamendo (MTG)~\cite{bogdanov2019mtg}.

\section{Proposed System}
\label{sec:proposed}


\begin{table*}[]
\centering
\sisetup{table-format=2.1,round-mode=places,round-precision=1,table-number-alignment = center,detect-weight=true,detect-inline-weight=math}
\begin{tabular}{lccSSSSSSSS}
\toprule
\multirow{2}{*}{\textbf{Method}} & \multirow{2}{*}{\textbf{Unsupervised?}} & \multicolumn{1}{c}{\multirow{2}{*}{\parbox{1.5cm}{\centering \textbf{Neural Network?}}}} & \multicolumn{3}{c}{MUSDB18~\cite{musdb18}} & \multicolumn{5}{c}{Slakh2100~\cite{manilow2019cutting}}               \\
\cmidrule(lr){4-6}\cmidrule(lr){7-11}
 &  &  & {Vocals} & {Bass} & {Drums} & {Bass} & {Drums} & {Guitar} & {Piano} & {Strings} \\
\midrule
Open-Unmix~\cite{stoter2019open}    &  & \ding{51}  & 14.972 & 11.850 & 11.598  & 9.753 & 14.395 & {\cellcolor{black!5} --} & {\cellcolor{black!5} --} & {\cellcolor{black!5} --}  \\
Demucs~\cite{defossez2019music}       &  & \ding{51}  & 15.494 & 13.147 & 12.659   & 10.759 & 15.533 & {\cellcolor{black!5} --} & {\cellcolor{black!5} --} & {\cellcolor{black!5} --}  \\
Cerberus~\cite{manilow2020simultaneous}  & & \ding{51} & {\cellcolor{black!5} --} &  8.301  & 7.474  & 10.843 & 15.370 & 10.181 & 10.520  & 12.4715  \\
\midrule
HPSS~\cite{fitzgerald2010harmonic} & \ding{51} & & {\cellcolor{black!5} --} & {\cellcolor{black!5} --}  & -0.069 & {\cellcolor{black!5} --} & 0.314 & {\cellcolor{black!5} --} & {\cellcolor{black!5} --} & {\cellcolor{black!5} --} \\
REPET-SIM~\cite{rafii2012music} & \ding{51}  &  & 7.816  & {\cellcolor{black!5} --} & {\cellcolor{black!5} --}  & {\cellcolor{black!5} --}  & {\cellcolor{black!5} --} & {\cellcolor{black!5} --} & {\cellcolor{black!5} --} & {\cellcolor{black!5} --} \\
\midrule
\textsc{TagBox} (Ours)  & \ding{51} & \ding{51} & 7.388 & 7.128 & 5.857 & 6.934 & 7.328  & 9.304 & 8.749 & 10.523  \\
\bottomrule
\end{tabular}
\caption{Comparison of source separation systems in terms of mean SDR improvement (dB) over the unprocessed mixture. Grey cells indicate that the system is unable to separate that source type. \textsc{TagBox} is the only system that is able to separate \textit{all} of the sources we test.}
\label{tab:sep_results}
\end{table*}

At the heart of our proposed system are two components: a pretrained generative music model (i.e., \textsc{Jukebox}) and a pretrained music tagging model. Because our system combines music taggers and \textsc{Jukebox}, we call our system \textsc{TagBox}. The core of the idea is simple, given an input audio clip, the generative model iteratively alters that input audio such that, when the altered audio is given to a tagger, the tagger's output increasingly matches a target set of tags that describe the desired set of sources.   An illustration of the proposed approach is shown in Figure~\ref{fig:diagram}. Algorithm 1 outlines the approach in pseudocode. We now describe the steps in our method.

We first create a target tag distribution $T_{target}$ by setting the tags that correspond to the desired instrument sources to $1$ (e.g., ``guitar'' or ``drums'') and all other tags to $0$. We then use $V_{encoder}$, the encoder portion of an autoencoder (in this case, the one in \textsc{Jukebox}), to produce an embedding $e$ from the input audio mixture $x$. This embedding is then decoded into a waveform $j$ by the decoder $V_{decoder}$.

Rather than pass $j$ directly to the $Tagger$, we use it as a mask on the input mixture. In this way, the embedding $e$ essentially determines what information must be \textit{removed} from the input mix to produce the desired source as defined by the tags.
For an input mixture waveform $x \in \mathbb{R}^{t}$ and a \textsc{Jukebox}-decoded waveform $j \in \mathbb{R}^{t}$, both with length $t$ samples, we convert both to a spectrogram representation, $X \in \mathbb{R}^{T \times F}$ and $J \in \mathbb{R}^{T \times F}$, with $T$ time frames and $F$ frequency bins. We then compute a real-valued mask, $\Bar{M} \in \mathbb{R}^{T \times F}$  as follows:

\begin{equation}
    \Bar{M} = \frac{|J|}{\max(|J|, |X|) + \varepsilon}
    \label{eq:mask}
\end{equation}

\noindent where $\max()$ is an element-wise max function between each time-frequency bin in a pair of spectrograms and a small epsilon, e.g $\varepsilon = 1e-8$, prevents division by zero. This mask $\Bar{M}$ is multiplied by the mixture spectrogram to get an estimate of the audio data that should be removed from the mix like $\Bar{S} = \Bar{M} \odot X$, where $\odot$ indicates element-wise multiplication. 
$\Bar{S}$ is then converted to a waveform of the source estimate $\Bar{s} \in \mathbb{R}^{t}$ using an inverse STFT. This waveform, $\Bar{s}$, is then put into the music tagger to determine the estimated tags. A binary cross-entropy loss  $\nabla L(T_{\Bar{S}}, T_{target};e)$ is computed between the estimate tags and the predetermined instrument tags. This loss is used to perform gradient ascent step in the \textsc{Jukebox} embedding space, where $\delta$ governs the step size. This approach is similar to adversarial example generation \cite{goodfellow2014explaining}, where the goal is also to optimize the input to produce a desired label.  
Because the mask made by the \textsc{Jukebox}-decoded audio determines what should be removed from the mix, the final estimate for a target source, $\hat{s}$, is the difference between the input mixture waveform $x$ and the final $\Bar{s}$ produced by gradient ascent. The final source estimate is therefore $s_{out} = x - \Bar{s}$.

We note that \textit{neither} the generative model \textit{nor} the music tagger were trained for source separation and that no additional training or alteration of the weights of either model happens at any point. These models were, however, trained on datasets with a wider range of audio than is typical for deep models trained specifically for source separation.

Our system is able to produce separation results for a larger set of sources than any previous deep learning system that we are aware of. This system is limited only by the tags of the music tagging system, of which there are 12 distinct instrument tags in MTG-Jamendo (MTG). MagnaTagATune (MTAT) has 31 tags that could be interpreted as instrument tags, although the tags conceptually overlap somewhat (e.g., MTAT contains distinct tags for ``vocals'', ``voice'', ``male vocals'', etc). Additionally, separating different source types does not require any changes to the system setup other than altering a set of predefined tags. Compare this to typical music separation networks like Open-Unmix~\cite{stoter2019open} which would require training a whole model for each new source or Demucs~\cite{defossez2019music} which would require altering the network architecture to add a new source output.

\begin{table*}[]
    \centering
    \sisetup{table-format=2.1,round-mode=places,round-precision=1,table-number-alignment = center,detect-weight=true,detect-inline-weight=math}
    \begin{tabular}{ccSSSSSSSS}
    \toprule
    \multicolumn{2}{c}{Tagger Settings} & \multicolumn{3}{c}{MUSDB18~\cite{musdb18}} & \multicolumn{5}{c}{Slakh~\cite{manilow2019cutting}} \\
    \cmidrule(lr){3-5}\cmidrule(lr){6-10}
    Dataset & Architecture & {Vox} & {Bass} & {Drums} & {Bass} & {Drums} & {Guitar} & {Piano} & {Strings} \\ \midrule
    \multirow{2}{*}{MTAT}   & FCN   & 7.943 & {\cellcolor{black!5} --}  & 5.687 & {\cellcolor{black!5} --}  & 7.317  & 9.551 & 8.596 & 10.412  \\
                        & HCNN & 6.641 & {\cellcolor{black!5} --}  & 4.955 & {\cellcolor{black!5} --}  & 6.506  & 8.788 & 7.322 & 8.484 \\
    \multirow{2}{*}{MTG-Jamendo}    & FCN  & 7.388 & 7.128 & 5.857 & 6.934 & 7.328  & 9.304 & 8.749 & 10.523  \\
                        & HCNN  & 6.758 & 6.744 & 5.777 & 6.690 & 7.280  & 8.332  & 8.077  & 8.980 \\
    \bottomrule
    \end{tabular}
    \caption{Comparison of using different pretrained, frozen taggers for gradient ascent with \textsc{TagBox} in terms of mean SDR improvement (dB) over the unprocessed mixture. 
    Note the MTAT taggers have no ``bass'' tag.}
    \label{tab:ablation}
\end{table*}

\section{Experimental Validation}

We conduct a series of experiments to validate our system, aimed at answering two questions. The first and main experiment is intended to compare the proposed system to existing systems, taking special care to try to understand \textsc{TagBox}'s ability to separate many types of sources. The second experiment is designed to determine how the choice of the pretrained, frozen Tagger model affects separation quality.

In our main experiment, we compare our system to existing systems on two established test sets for source separation, namely MUSDB18~\cite{musdb18} and Slakh2100~\cite{manilow2019cutting}. In this experiment we compare our proposed system against recent deep learning-based supervised separation systems as well unsupervised separations based on musical priors. We compare our system on a wide variety of source types across both of these datasets.

The first dataset we examine is MUSDB18. MUSDB18 contains 150 mixtures and corresponding sources from real live recording sessions, 100 of these are reserved for training and the remaining 50 are used for testing. For this experiment, we exclude MUSDB18's ``other'' source because it could map to many possible tags using \textsc{TagBox}. The supervised systems that we compare against, namely Open-Unmix~\cite{stoter2019open} and Demucs~\cite{defossez2019music}, are trained using the MUSDB18 training set. Contrast this to the unsupervised systems we test, HPSS~\cite{fitzgerald2010harmonic} and REPET-SIM~\cite{rafii2012music}, which are run on the test set without any training. Our proposed system falls into this second camp; it is also unsupervised and therefore does not have a training phase, ignoring the MUSDB18 training set. 

The main experiment also uses the Slakh2100~\cite{manilow2019cutting} dataset. Slakh2100 contains 2100 mixtures with corresponding sources that were synthesized using professional-grade sample-based synthesis engines. We chose 50 songs from the test set 
to evaluate on. We chose songs that have source data for following five source types: bass, drums, guitar, piano, strings. We select mixes where all 5 sources are active, and we say a source is active if it has 100 or more note onsets throughout the entirety of the song, as determined by the corresponding MIDI data.
We create mixes by instantaneously mixing together the sources and use these mixtures as input to the systems. With this setup we compare against Cerberus~\cite{manilow2020simultaneous}, which was trained to separate these five instruments, specifically.

For \textsc{TagBox}, we use a pretrained FCN~\cite{choi2016automatic} tagger trained on the MagnaTagATune (MTAT)~\cite{law2009evaluation} dataset. We run gradient ascent with a learning rate of 5.0 using the Adam optimizer for 10 steps (in the interest of brevity), and use a spectrogram with 1024 FFT bins for the mask. Additionally, we use the ``foreground'' from REPET-SIM as the vocals estimate, following prior work~\cite{rafii2012music}, and use the ``percussion'' output from HPSS as the drums estimate. We omit the other source outputs of these systems because they are ill-defined (e.g., HPSS's ``harmonic'' could be many possible sources).

In the second experiment, we compare four different configurations of our proposed system, varying the architecture and training data of the music tagger. We look at the FCN~\cite{choi2016automatic} and HarmonicCNN~\cite{won2020data} architectures, trained either MagnaTagATune (MTAT)~\cite{law2009evaluation} or MTG-Jamendo~\cite{bogdanov2019mtg}. We use the same learning rate and number of steps as the previous experiment.


We evaluate the outcome of our experiments using the source-to-distortion ratio improvement (SDRi) over the unprocessed mixture~\cite{vincent2006performance} using the \textit{museval} toolbox~\cite{SiSEC18}.



\section{Results and Discussion}
\label{sec:results}

Table~\ref{tab:sep_results} shows the results of our main experiment. 
In terms of SDRi, our system is better than or competitive with both of the hand-designed unsupervised algorithms that we test against, HPSS and REPET-SIM. 
Additionally, while our system does not show as good of performance as the purpose-built supervised separation systems (i.e., Open-Unmix, Demucs, and Cerberus), it still shows a considerable SDRi boost for all sources that we test. Importantly, our system is able to boost performance over a wider array of source types than any other system we compare against.

The results from our second experiment are shown in Table~\ref{tab:ablation}. Of the two architectures we test, using FCN always produces better separation results. Interestingly, the opposite trend was observed when the taggers were evaluated for music tagging performance by Won et. al.~\cite{won2020eval}: HCNN was among the top performing systems and FCN was towards the bottom of the pack. 

In many cases, \textsc{TagBox} can leave much to be desired perceptually; in most cases its separation performance is not up to the same level as the purpose-built separation systems we compare against. However, when listening to the output, there is no doubt that \textsc{TagBox} is able to separate the desired source, despite apparent artifacts. We have informally noticed a few tricks for better perceptual performance, like using multiple FFT sizes when making the masks (\textit{à la} a multi-scale spectral loss) and doing gradient ascent for 100 steps. These perceptual tricks were however not reflected in the SDR evaluation numbers. Furthermore, because producing each output example requires its own gradient ascent, adding more steps increases the computation time linearly, which can be a costly process when run on an entire dataset. However, this might be tolerable for musicians needing a flexible source separation solution on a single song.

There are also a few other variants of the \textsc{TagBox} setup that can lead to fun and unexpected creative results. In the first case, we remove the masking step and allow \textsc{TagBox} to create audio freely, without the constraint of having to only remove information from the mix. With this setup, \textsc{TagBox} performs a kind of style transfer, mapping certain features of the audio to the desired tag. In one example, a mixture had a singer and we selected the ``guitar'' tag. \textsc{TagBox} made the resultant audio sound like a guitar was performing the melody. Additionally, another variant involves selecting non-instrument tags, like genre tags, and optimizing those.


What we find most impressive is that neither \textsc{Jukebox} nor the music taggers were trained for source separation. Furthermore, the weights of both networks do not change during the gradient ascent process; only the location of the audio in the \textsc{Jukebox} embedding space changes. The combination of \textsc{Jukebox} and the taggers have seen up to 1.25 million songs
and combined these systems are able to leverage their shared priors about music and musical sources to isolate individual musical sources in a mixture. We believe that these priors could be leveraged in many ways to overcome the data scarcity problems endemic to many MIR tasks, as has already been investigated to great effect by Castellon et. al~\cite{castellon2021codified}. We are excited about future explorations in this area.



\section{Conclusion}
\label{sec:conclusion}

In this paper, we have proposed a method for unsupervised source separation by combining pretrained models, called \textsc{TagBox}. We use pretrained music taggers to do gradient ascent in the embedding space of OpenAI's \textsc{Jukebox} with the goal of maximizing a pre-defined tag corresponding to the source we want to separate. The output of \textsc{Jukebox} is used as a mask on the input audio before being sent to the tagger, which ensures that \textsc{Jukebox} does not generate new, unseen data that is not present in the input mixture. Importantly, neither the tagger nor \textsc{Jukebox} have been trained for source separation and the weights of both models remain fixed during the gradient ascent process. We demonstrate results showing that our system is able to separate a wider variety of source types than many recent purpose-built, supervised separation systems. We are excited by the promise that pretrained systems hold for the future of MIR and source separation research.

\section{Acknowledgements}
\label{sec:ack}

The authors would like to thank Ian Simon, Sander Dieleman, Jesse Engel, and Curtis Hawthorne for their fruitful conversations about this work. Additionally, we would like to thank the creators of \textsc{Jukebox} for help with their codebase: Prafulla Dhariwal, Heewoo Jun, Christine Payne, Jong Wook Kim, Alec Radford, and Ilya Sutskever.


\bibliographystyle{IEEEbib}
\bibliography{strings,refs}

\end{document}